\documentclass{aa}

\usepackage{epsfig}
\usepackage{graphicx}
\newcommand{\um}{\,$\mu$m}

\begin{document}

\title{Broad-line radio galaxies: old and feeble?}
\author{Ilse van Bemmel\inst{}, Peter Barthel\inst{}}
\offprints{bemmel@astro.rug.nl}

\institute{Kapteyn Astronomical Institute, P.O.~Box 800,
NL--9700\,AV Groningen, The Netherlands} 

\date{Received date; accepted date}

\abstract{Far-infrared photometry of broad-line radio galaxies shows
this class of AGN to consist of many hot and some cool infrared
emitters, with peaks in their spectral energy distributions around
25\,$\mu$m or longward of 60\,$\mu$m, respectively.  Quantitative
analysis indicates that this distribution relates to a substantial
dispersion in the strength of the cool dust component: broad-line radio
galaxies are relatively poor in large-scale dust. Possibly they have
undergone a different merger evolution, or are relatively old AGN.
}

\maketitle

\section{Introduction}

Broad-line radio galaxies (BLRGs) were discovered and classified in the
1960's, as part of radio source identification programs (e.g., Matthews
et al.  1964).  Having optically visible host galaxies and optical
spectra with broad recombination lines, they exhibit characteristics of
both the narrow-line radio galaxies (NLRGs) and radio-loud quasars
(QSRs).  Some of these BLRGs were identified with galaxies having a
"brilliant starlike nucleus" (Matthews et al.  1964), the so-called
N-galaxies (Morgan 1958).  Their strong nuclear optical emission was
subsequently discovered to experience reddening (Sandage 1973).  It now
appears that many BLRGs are hosted by such N-type galaxies (e.g., Hes et
al.  1995).  A thermal, accretion disk origin for their bright nuclear
continuum has recently been established (e.g.  Chiaberge et al.  2000). 
Unified models for radio-loud AGN postulate the presence of an opaque
dust distribution (torus) in the radio axis equatorial plane,
surrounding the continuum source and broad emission line region (Barthel
1989, Urry \& Padovani 1995).  Within this framework BLRGs can be
considered objects for which the line of sight is close to the torus
edge.  Thus the broad-line region is directly seen, but the central
engine is still partially hidden from view. \\
\indent Consistent with this picture, Cohen et al.  (1999) found the optical
spectra of several BLRGs to be a super-position of a reddened quasar
spectrum and polarized (reflected) quasar emission.  The reddening
appears to increase for objects in which the polarization is higher,
which can be interpreted as an increase in inclination.  Using radio
imaging and polarization data to address the aspect geometry,
Dennett-Thorpe et al.  (2000) found the class of BLRGs to consist of two
groups.  One contains the objects which are powerful mis-aligned quasars
in which the central engine is highly reddened.  These objects are
observed at relatively large inclination angles.  The other consists of
nearby, lower luminosity quasars, for which their proximity and low AGN
luminosity combine to yield the observed broad-line radio galaxy.  Like
QSRs, these are observed at relatively small inclination angles.  Prime
example is 3C\,111, for which superluminal behaviour in a one-sided
radio jet -- a typical QSR phenomenon -- has been measured (Goetz et al. 
1987). \\
\indent A third way of addressing the aspect angle towards BLRGs is using
infrared (IR) observations.  The dust torus will absorb and re-emit the
soft X-rays and UV emission of the AGN, and the emission will be
anisotropic at wavelengths where the dust is optically thick.  Radiative
transfer models show that there is indeed a clear aspect angle
dependence of the spectral energy distribution (SED) in the near- and
mid-IR (Granato \& Danese 1994, Pier \& Krolik 1992).  Consistent with the 
model
predictions, QSRs are generally found to be stronger and somewhat warmer
in the near- and mid-IR than NLRGs.  This difference may well persist to
wavelengths $\approx 60\,\mu$m (van Bemmel et al.  2000).  If BLRGs are
mis-aligned and low-luminosity QSRs their IR SEDs are expected to be
intermediate between those of QSRs and NLRGs. \\
\indent The SED far-infrared (FIR) peak for QSRs and NLRGs is observed between
60 and 100\,$\mu$m.  However, IRAS data revealed several BLRG to display
considerably warmer IR emission, indicating the presence of much hotter
dust than in QSRs/NLRGs.  Following the measurement of a pronounced
25\,$\mu$m peak in 3C\,390.3 (Miley et al.  1984), more such cases were
reported by Hes et al.  (1995).  In this paper we will study the FIR SED
of the BLRG class, in particular the nature of the short wavelength
peak.  We will refer to the objects with a 25\,$\mu$m peak as {\em hot},
while the others will be called {\em cool}. 

\begin{figure}[!t]
\resizebox{8.7cm}{!}{\hspace{-5mm}\includegraphics{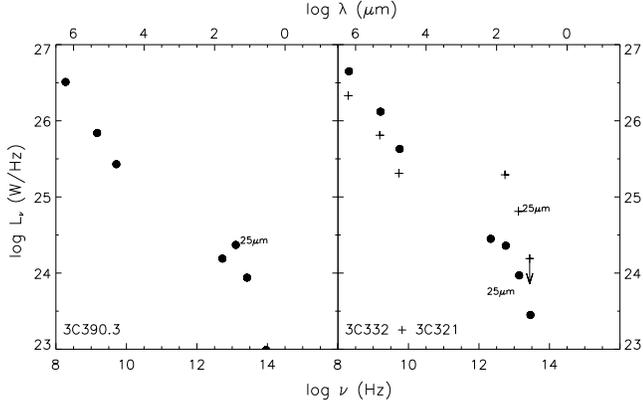}}
\hfill
\parbox[b]{8.7cm}{
\caption{\label{seds}
Left panel: the restframe SED for the BLRG 3C\,390.3
(hot BLRG). Right panel: the restframe SEDs for BLRG 3C\,332 
(filled circles) and NLRG 3C\,321 (pluses), both cool objects.
The 25\,$\mu$m points are marked in the plots. Errors 
are comparable to the size of the 
plot symbols.}}
\vspace{-2mm}
\end{figure}

\section{Observations and data reduction}

We carried out an ISO (Kessler et al.  1996) program, using ISOPHOT
(Lemke et al.  1996), to obtain IR photometry for eight double-lobed
BLRGs.  The sample consisted of 3C\,17, 3C\,33.1, 3C\,34, 3C\,59,
3C\,61.1, 3C\,330, 3C\,332 and 3C\,381, and was selected from Hes et al. 
(1995).  Observations were made at 12, 25, 60, 90 and 160\,$\mu$m, using
the P1, P2, C1 and C2 detectors.  Raster mode observing was applied, to
avoid the known problems with chopping.  The reduction of the C1 and C2
data followed the description in Van Bemmel et al.  (2000).  The
reduction of the P1 and P2 data is similar, the loss of redundancy is
compensated by higher quality detectors, resulting in better data for
the P-detectors.  Unfortunately, 3C\,34 was not detected at any
wavelength, 3C\,330 was not observed at 25\um, and 3C\,17 was not
observed at both 25 and 60\um, which lead us to remove them from the
sample.  3C\,59 was rejected because of confusion with a nearby Seyfert
galaxy.  For two of the remaining objects the classification as a BLRG
appears dubious.  3C\,61.1 had been erroneously identified with a nearby
radio-quiet QSO -- hence its broad line spectrum; on the basis of a new
high-resolution radio image it now appears to be an NLRG (Dennett-Thorpe
et al.  2000).  3C\,381 is reported to have broad Balmer line wings
(Grandi \& Osterbrock 1978), but no literature spectra are available to
confirm this.  Dennett-Thorpe et al.  (2000) conclude that 3C\,381 is
more likely to be an NLRG. \\
\begin{figure}[!t]
\centerline{\epsfysize=8.truecm
\resizebox{4.3cm}{!}{\epsfbox{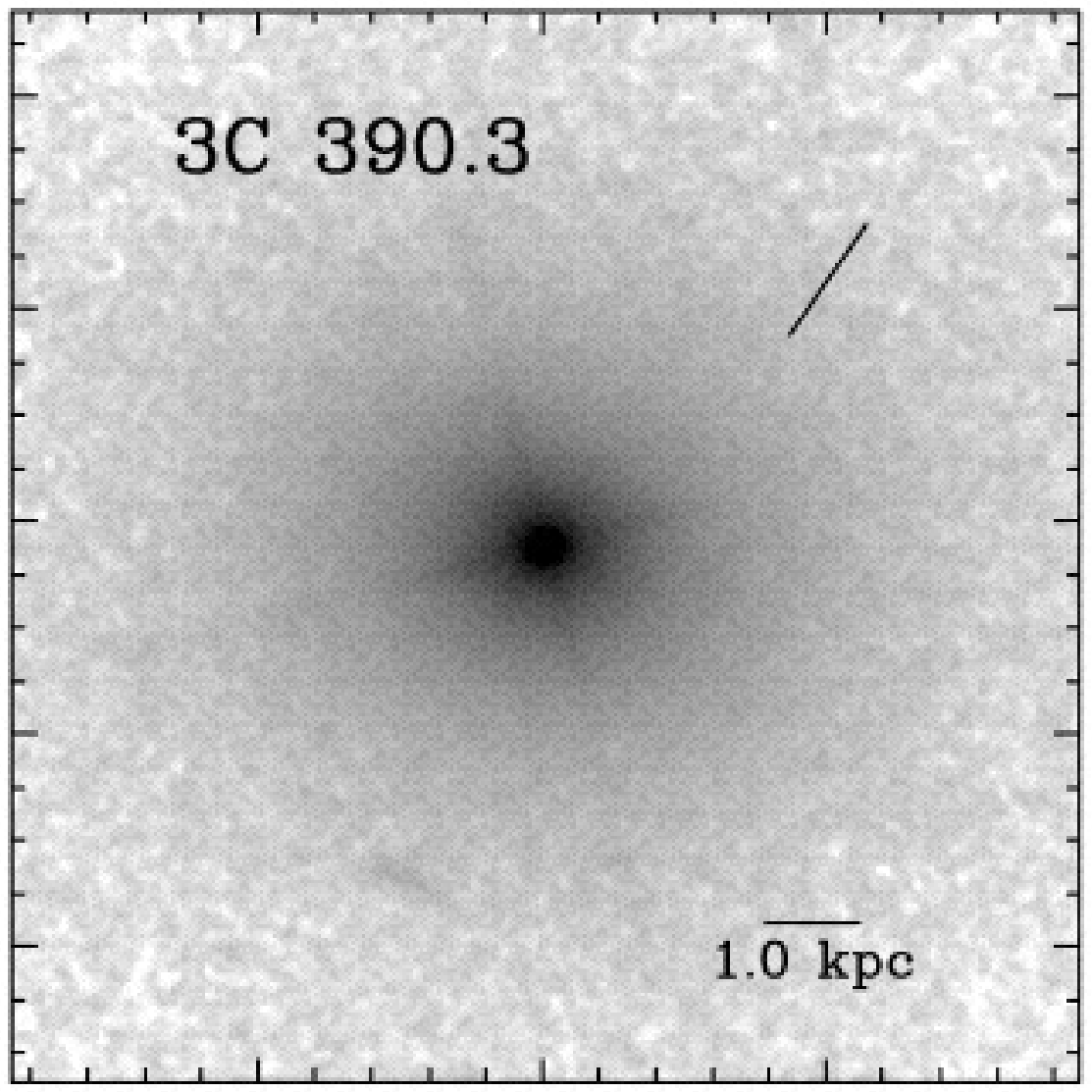}}
\resizebox{4.3cm}{!}{\epsfbox{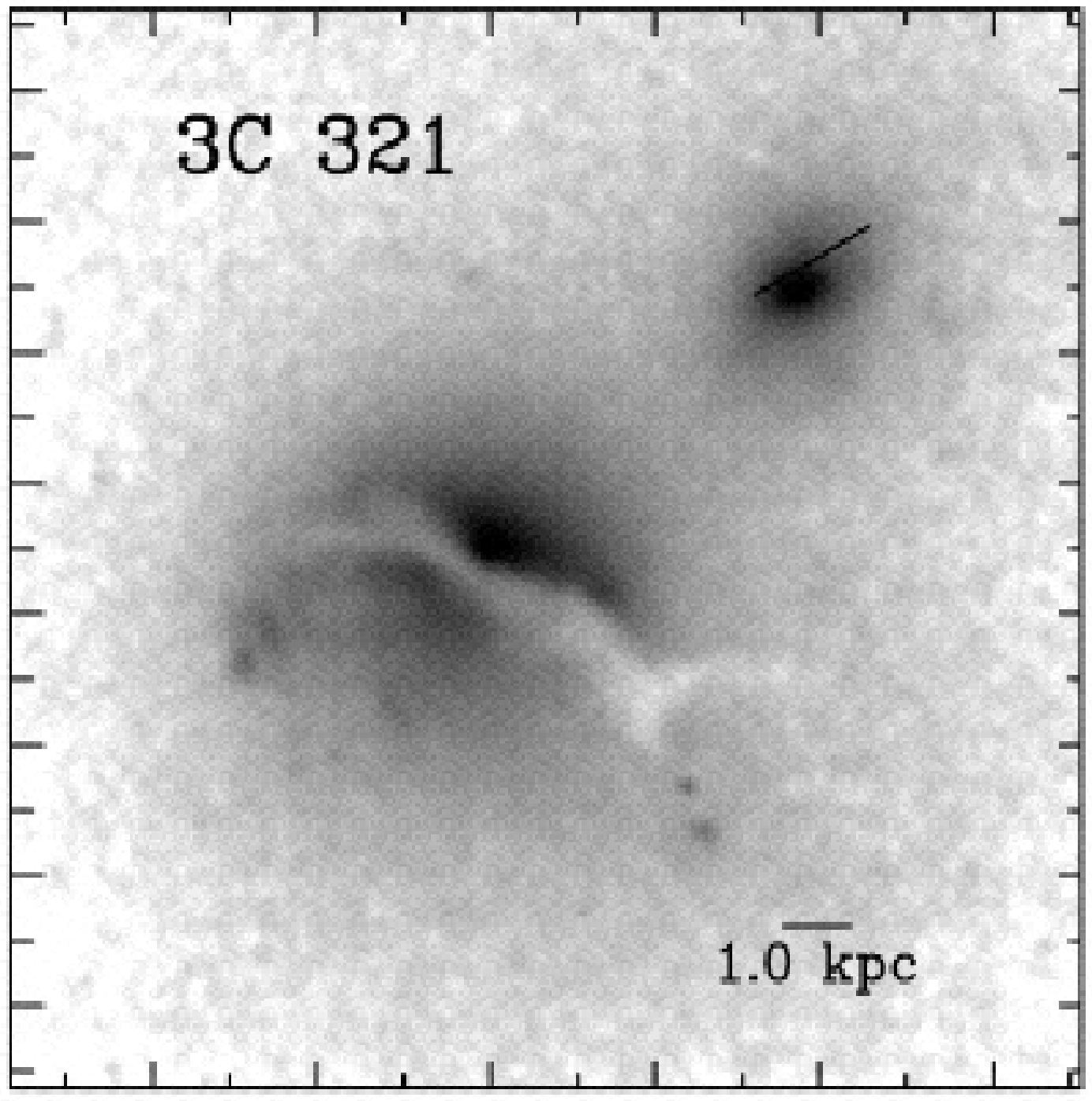}}}
\parbox[b]{8.7cm}{
	\caption{\label{images}
	Optical HST images of 3C\,390.3 (hot BLRG) and 3C\,321 (cool NLRG). 
	The black lines indicate the radio axis and the image scale
	respectively (de Koff et al. 1996).}
}
\vspace{-2mm}
\end{figure}
\indent We extracted ISOPHOT archive data for BLRG 3C\,109, and added five 3C
BLRGs with IRAS detections at 25 and 60\,$\mu$m, enlarging the BLRG
database to a total of 8 bona fide objects, in the redshift range $0 < z
< 0.3$.  3C\,98 was included, having a 25\um\ detection in excess of its
60\um\ upper limit.  3C\,234 and 3C\,381 were added, in spite of their
dubious BLRG nature (Cohen et al.  1999, Dennett-Thorpe et al.  2000). 
Lacking evidence that they are in fact NLRG, they will be treated as a
BLRG, bringing the total to 10 objects. \\
\indent It appears that 7 out of 10 BLRGs display a peak around 
25\,$\mu$m, with
a positive spectral index\footnote{We assume $F_{\nu} \propto
\nu^{\alpha}$} between 60 and 25\um.  Three BLRGs show a small negative
$\alpha_{25}^{60}$ value, which is still in marked contrast to NLRGs,
displaying typical values $\alpha_{25}^{60} \approx -1.2$ (Hes et al. 
1995).  In order to compare objects of positive $\alpha_{25}^{60}$ with
objects of negative $\alpha_{25}^{60}$, we supplement the BLRG sample
with five 3C NLRGs of comparable redshift and radio luminosity.  These
NLRGs are obviously selected to have known 25 and 60\,$\mu$m flux
densities from either IRAS (Golombek et al.  1988) or ISO (3C\,61.1 --
our measurements), and to have double-lobed radio morphologies.\\
\indent Subsequently, the total sample of fifteen powerful FR\,II 
radio galaxies
is divided into two sub-samples of hot and cool objects, depending on
their value of $\alpha_{25}^{60}$, and regardless of their optical
spectral classification.  The hot sub-sample is defined as having a
positive or flat spectral index, the cool subset has a negative spectral
index.  Typical SEDs for hot and cool BLRGs are shown in
Fig.~\ref{seds}.  To complement the IR SEDs and confirm the thermal
nature of the IR emission, radio data have been extracted from NED and
the literature.  The two samples are listed in Table~\ref{lum}.  The
luminosities have errors of 1\% or less, and take redshift and
K-correction effects into account. 

\begin{table*}[!ht]
\begin{center}  
\begin{tabular}{|lclcccc|lclcccc|}
\hline
\multicolumn{7}{|c|}{\bf Hot (peak around 25\um)} &
 \multicolumn{7}{c|}{\bf Cool (peak longward of 60\um)} \\
\hline
Name& $z$ &  ID & Type & $P_{25}$ & $P_{60}$& $P_{178}$ & Name& $z$ & ID & Type & $P_{25}$&  $P_{60}$& $P_{178}$ \\
\hline
3C\,33.1& 0.181& B  & --   & 24.41 & 24.36 & 26.93 & 3C\,61.1& 0.184& N   & $+$ &23.97 & 24.59 & 27.39 \\
3C\,98  & 0.030& B  & --   & 23.19 &$<$23.15&25.92 & 3C\,79  & 0.256& N   & $+$& 24.95 & 25.40 & 27.52 \\
3C\,109 & 0.306& B  & $+$  & 25.78 & 25.36 & 27.60 & 3C\,111 & 0.049& B   & -- & 24.01 & 24.16 & 26.20 \\
3C\,234 & 0.185& B? & $+$  & 25.29 & 25.14 & 27.38 & 3C\,321 & 0.096& N   & $+$& 24.84 & 25.30 & 26.31 \\
3C\,382 & 0.058& B  & --   & 23.78 & 23.82 & 26.11 & 3C\,327 & 0.104& N   & $+$& 24.80 & 25.16 & 26.96 \\
3C\,390.3&0.056& B  & --   & 24.36 & 24.15 & 26.50 & 3C\,332 & 0.152& B   & $+$& 23.99 & 24.36 & 26.63 \\
3C\,445 & 0.056& B  & --   & 24.32 & 24.21 & 26.16 & 3C\,381 & 0.161& B?  &  ? & 24.31 & 24.48 & 26.92 \\
        &      &    &   &       &       &          & 3C\,403 & 0.059& N   & -- & 24.17 & 24.47 & 26.27 \\
\hline
{Average}   & 0.124   & & &24.45 &$<$24.31& 26.65& {Average}& 0.133 & & & 24.38   & 24.74 & 26.78\\
\hline
\end{tabular}

\parbox[b]{17.8cm}{\vspace{2mm}   
\caption{\label{lum}
Luminosity densities in W\,Hz$^{-1}$ at 25\,$\mu$m, 60\,$\mu$m and
178\,MHz for the hot and cool samples. Average luminosities for the hot and
cool samples are given for comparison. The third column lists the nature of 
the object: B for BLRG and N for NLRG. 3C\,234 and 3C\,381 are marked
B? in accordance with findings of Cohen et al. (1999) and Dennett-Thorpe 
et al. (2000). The fourth column lists the optical
appearance: $+$ for a disturbed host galaxy, -- for a normal elliptical 
appearance, and ? if indecisive.  
We used $H_0$\,=\,75\,km\,s$^{-1}$\,Mpc$^{-1}$
and $k=\Lambda$\,=\,0 (i.e. $q_0 = \frac{1}{2}$).}
}
\vspace{-9mm}
\end{center}
\end{table*}
\normalsize

\section{Results and analysis}

From Table~\ref{lum} we see that both sub-samples are dominated by a
different class: there are no hot NLRGs, and hardly any cool BLRGs. 
This is entirely in agreement with findings of Yates \& Longair (1989),
Impey \& Gregorini (1993), Heckman et al.  (1994) and Hes et al. 
(1995), and may provide an important clue to the nature of BLRGs. \\
\indent Although incomplete, literature data did not reveal any differences
between X-ray, optical, near-IR and radio properties of the sub-samples. 
With reference to Table~\ref{lum}, we however establish a significant
difference between the 60\um\ luminosities of the hot and the cool
sample, while their 25\um\ and 178\,MHz luminosities are comparable. 
Thus, the hot broad-line objects must owe their classification as hot to
their relative weakness at 60\um.  Assuming the IR SED is dominated by
thermal emission, this implies the absence of cool ($\sim$\,50\,K) dust
in the hot objects.  Hot ($\sim$ 150\,K) dust is present in all objects
at a similar level, but it dominates the SED only in the hot sample. 
The visibility of broad emission lines must somehow be connected to 
relatively weak cool dust emission in this sample, and we proceed by
addressing the origin of this cool dust. \\
\indent There are two processes that heat dust up to $\sim$\,50\,K in radio
galaxies, namely the AGN (e.g., Hes et al.  1995) and star-formation. 
Detailed examination of the FIR SEDs of Seyfert galaxies (Rodr\'{\i}guez
Espinosa et al.  1996), has indicated that the AGN strength is primarily
reflected in the strength of the hot (T$\sim$\,150\,K) dust, i.e., the
25\um\ power.  Prieto (2001) shows that the hot dust emission is
correlated with the hard X-ray flux, whereas the cooler emission is not. 
This is consistent with our finding that 178\,MHz and 25\um\
luminosities, both reflecting the AGN power, are comparable between the
hot and cool sample.  We therefore exclude the AGN as the dominant
heating mechanism.  The FIR SED of the cool sample is comparable to that
of dusty starburst galaxies, which generally have a peak in their SED
around 60\,$\mu$m (e.g.  Sanders \& Mirabel 1996).  We therefore
hypothesize that star-formation dominates the heating of cool dust to
50\,K, in which case the hot objects must have significantly less
star-formation. \\
\indent In the absence of H$\alpha$ and/or UV data, we test this hypothesis
using broad-band imaging data from the HST snapshot survey of the 3CR
sample (de Koff et al.  1996).  A three person blind test was done to
classify the sample galaxies as non-interacting/dust-free,
interacting/dust-rich, or unclear: the results are summarized in
Table~\ref{lum}.  A clear relation is found between the visual
appearance of the host galaxy 
and the shape of
its IR SED.  Hot BLRGs are predominantly normal, relaxed, dust-free
elliptical galaxies, usually residing in sparse environments.  The cool
objects, however, display dust lanes, sometimes have close companions,
and often show signs of interaction.  The ubiquitous dust appears
extended over tens of kpc and is mostly oriented perpendicular to the
radio axis.  In Fig.~\ref{images} two typical images for each sample are
shown. Since the only disturbed cool BLRG 3C\,332 has no clear
dust lane, but is simply a relaxed elliptical with a nearby
companion, we show cool NLRG 3C\,321 for a more representative
comparison. Both SEDs are included in Fig.~\ref{seds}. This striking
difference in visual appearance indicates that many BLRGs are hosted by
relaxed, dust- and gas-deficient galaxies, and provides, in addition, a 
natural explanation for the observed FIR SED difference. 

\section{Discussion}

In line with trends reported by Hardcastle et al.  (1998), some BLRGs
such as superluminal, cool 3C\,111 and optically variable, hot 
3C\,390.3 are undoubtedly low-power QSRs, observed at small inclination
angles.
Cohen et al.  (1999) and Dennett-Thorpe et al.  (2000) however
point out that other BLRGs must be identified with critically inclined
QSRs, i.e., radio sources observed close to or through their torus
boundary.  Since the dust content of a galaxy is unrelated to the
orientation of its central radio source, a sizeable fraction of the BLRG
population must constitute a class of its own. As a consequence,
the clear line of sight towards the broad-line region in these hot BLRGs 
must be related to their dust- and gas-deficient host galaxies,
either on large or on small scale.
In the first case this implies that the obscuration of the broad-line
region, generally assumed to occur in the circumnuclear torus (on sub-kpc
scale), can also occur
on much larger scales. Similar suggestions have been made for 
Seyfert galaxies (Malkan et al. 1998). However, there is no reason why
the lack of cold dust on large scales does not extend into the nuclear 
regions, and thus the geometry of a central torus can be affected.
Most likely, the torus is not solid, but exists of overlapping dense 
cool clouds. In the absence of a large supply of cold dust, these clouds 
may be smaller and less numerous, yielding low-extinction sight lines 
toward the central regions.\\
\indent 
As described above and has been reported elsewhere in much more
detail (e.g., Heckman et al.  1986, Aretxaga et al.  2001),
host galaxies of powerful radio sources frequently display
signatures of ongoing or recent star-formation, which is possibly
connected to the onset of nuclear activity (O'Dea et al.  2001).  Also
several Seyfert galaxies and QSO hosts display such signatures, and age
dating of their stellar population (e.g., Gonz\'alez Delgado et al. 
2001, Canalizo \& Stockton 2001) is strongly suggestive of an earlier,
interaction-induced starburst phase, which produces and drives the
fueling of the nuclear activity (Sanders et al.  1988, Barthel 2001). 
We observe no signs of this connection in the hot BLRGs.
Nuclear activity and starbursts are generally assumed to be triggered 
by a major
merger, and there are two ways to generate a relatively dust free
radio galaxy. The first possibility is that such a radio galaxy forms from 
a single major dust-rich merger, the dust being removed during and
immediately after the initial starburst phase.
This scenario would be consistent with comparable 
AGN power and host galaxy luminosities for cool and hot objects. 
However, the dynamical
timescales of large scale dust disks are of order 10$^9$ years, about an 
order of magnitude larger than the maximum age of the radio source.
A second possibility is that the hot BLRGs have been formed out of 
multiple dust-poor mergers. This results in comparable host galaxies 
and black hole masses, and naturally explains the lack of cold gas and 
dust in the hot objects (Haehnelt \& Kauffmann 2000).\\
\indent Finally, if the onset of nuclear activity is preceded by a
starburst phase, the ceasing of star-formation might be connected 
to the final evolutionary stages of the
AGN. When all dust and gas has been dissipated, it is no longer possible
to power star-formation and the active nucleus, and to sustain an
obscuring torus. We note with interest that this scenario is consistent 
with the fact that the hot BLRGs in our sample are all relatively large 
double-lobed radio sources.\\
\indent We plan to investigate these models using spectropolarimetric
observations. The spectra of the underlying host galaxies will provide 
clues to the stellar population, and thus information on the merger history
and age of the objects. Measurements of the cool gas content 
will also be relevant.

\section{Conclusions}

We have shown that hot objects, residing in dust-poor host galaxies,
dominate the BLRG population. Their 
optical appearance shows little, if any, extended dust features, and 
their FIR SED is lacking emission from cool, star-formation related dust.
This extended dust might contribute to the obscuration of the broad-line
region in NLRGs. The lack of it in hot BLRGs can be explained either
by a different merger history, or by their evolutionary stage. 

\acknowledgements

The authors wish to thank Marshall Cohen and referee Chris O'Dea
for detailed comments. Thanks
are also due to Jane Dennett-Thorpe and Ronald Hes for their 
initial involvement in these BLRG investigations.\\
The NASA/IPAC Extragalactic Database (NED) is operated by the
Jet Propulsion Laboratory, California Institute of Technology, under 
contract with the National Aeronautics and Space Administration. 
ISO is an ESA project with instruments funded by ESA member states 
(especially the PI countries: France, Germany, the Netherlands and 
the United Kingdom) and with participation of ISAS and NASA.

\end{document}